\documentclass[aps,prd,superscriptaddress,preprint,tightenlines,nofootinbib,floatfix]{revtex4}

\usepackage{graphicx}
\usepackage{dcolumn}
\usepackage{bm}
\usepackage{epsfig}
\newcommand{\Dsstarp}{D_s^{\star +}}

\newcommand{\Dsm}{D_s^-}

\newcommand{\Dp}{D^+}

\newcommand{\Kp}{K^+}
\newcommand{\Km}{K^-}
\newcommand{\pim}{\pi ^-}
\newcommand{\Dstar}{D^{\star}}
\newcommand{\Ks}{K^0_S}

\newcommand{\Kstarz}{K^{\star 0}}
\newcommand{\Ds}{D_s}
\newcommand{\Dz}{D^0}
\newcommand{\Dm}{D^-}
\newcommand{\epm}{e^+e^-}
\newcommand{\Dsp}{D_s^+}
\newcommand{\Dsstarm}{D_s^{\star -}}
\newcommand{\Dstarp}{D^{\star +}}
\newcommand{\Dsstar}{D_s^{\star}}
\newcommand{\piz}{\pi^{0}}
\newcommand{\etap}{\eta ^\prime}
\newcommand{\pip}{\pi ^+}
\newcommand{\rhom}{\rho ^-}
\newcommand{\Kstar}{K^{\star0}}
\newcommand{\MMtwos}{\text{MM}^{\star 2}}

\newcommand{\MDs}{M_{D_s}}
\newcommand{\chisq}{\rm{\chi^{2}}}

\newcommand{\etappi}{$\eta ^{\prime}(\eta \pi^+ \pi^-) \pi^{-}$}
\newcommand{\kkpipiz}{$K^{-}K^{+}\pi^{-}\pi^{0}$}

\newcommand{\ksks}{$K^{\star -}K^{\star 0}$}

\newcommand{\epirho}{$\pi^-\eta^{\prime}(\rho\gamma)$}

\newcommand{\kpipim}{$K^{+}\pi^{-}\pi^-$}
\newcommand{\kpipipizm}{$K^{+}\pi^{-}\pi^- \pi^{0}$}
\newcommand{\kzpim}{$K^{0}_{S}\pi^{-}$}
\newcommand{\kzpipipim}{$K^{0}_{S}\pi^{-}\pi^+ \pi^-$}
\newcommand{\kzpipizm}{$K^{0}_{S}\pi^{-}\pi^0$}
\newcommand{\kkpipm}{$K^{+}K^{-}\pi^{-}$}

\newcommand{\phienup}{$D_s^{+} \to \phi e^{+} \nu_e$}
\newcommand{\etaenup}{$D_s^{+} \to\eta e^{+} \nu_e$}
\newcommand{\etapenup}{$D_s^{+} \to\eta^{\prime} e^{+} \nu_e$}
\newcommand{\fzeenup}{$D_s^{+} \to f_{0} e^{+} \nu_e$}
\newcommand{\kstenup}{$D_s^{+} \to K^{\star 0} e^{+} \nu_e$}
\newcommand{\kzeenup}{$D_s^{+} \to K^{0} e^{+} \nu_e$}

\begin{document}



\title{Absolute Branching Fraction  Measurements for\\
        Exclusive  $D_s$ Semileptonic Decays}

\author{J.~Yelton}
\affiliation{University of Florida, Gainesville, Florida 32611,
USA}
\author{P.~Rubin}
\affiliation{George Mason University, Fairfax, Virginia 22030,
USA}
\author{N.~Lowrey}
\author{S.~Mehrabyan}
\author{M.~Selen}
\author{J.~Wiss}
\affiliation{University of Illinois, Urbana-Champaign, Illinois
61801, USA}
\author{R.~E.~Mitchell}
\author{M.~R.~Shepherd}
\affiliation{Indiana University, Bloomington, Indiana 47405, USA }
\author{D.~Besson}
\affiliation{University of Kansas, Lawrence, Kansas 66045, USA}
\author{T.~K.~Pedlar}
\affiliation{Luther College, Decorah, Iowa 52101, USA}
% ...
\collaboration{CLEO Collaboration}
\noaffiliation


\date{\today}

\begin{abstract}
We measure the absolute branching fractions of  $\Ds$ semileptonic
decays where the hadron in the final state is one of $\phi$,
$\eta$, $\etap$, $\Ks$, $\Kstar$, and $f_0$, using $2.8 \times
10^5$ $\epm \to \Ds\Dsstar$ decays collected in the CLEO-c
detector at a center-of-mass energy close to 4170 MeV. We obtain
${\cal B}(\Dsp\to \phi e^+ \nu _e) =(2.29\pm 0.37 \pm 0.11)$\%,
${\cal B}(\Dsp\to \eta e^+ \nu _e) =(2.48 \pm 0.29 \pm 0.13)$\%,
${\cal B}(\Dsp\to \etap e^+ \nu _e) =(0.91 \pm 0.33 \pm 0.05)$\%,
where the first uncertainties are statistical and the second are
systematic. We also obtain ${\cal B}(\Dsp\to K^0 e^+ \nu _e)
=(0.37\pm 0.10 \pm 0.02)$\%, and ${\cal B}(\Dsp\to K^{\star 0} e^+
\nu _e) =(0.18 \pm 0.07 \pm 0.01)$\%, which are the first
measurements of Cabibbo suppressed exclusive $\Ds$ semileptonic
decays, and, ${\cal B}(\Dsp\to f_0 e^+ \nu _e) \times {\cal B}(f_0
\to \pip\pim) =(0.13\pm 0.04 \pm 0.01)$\%. This is the first
absolute product branching fraction determination for a semileptonic decay including a scalar meson
in the final state.
\end{abstract}

 \maketitle


The study of $\Ds$ semileptonic decays provides interesting
information on several aspects of heavy quark decays. First of
all, the total semileptonic width provides discrimination between
different theoretical evaluations of hadronic matrix elements
affecting charm semileptonic decays. The Operator Product
Expansion (OPE) predicts that all the charmed mesons have the same
semileptonic width, modulo non-factorizable corrections
\cite{voloshin}. The ISGW2 form factor model \cite{ISGW2} predicts
a difference between the $D$ and $\Ds$ inclusive rates, as the
spectator quark masses $m_u$ and $m_s$ differ on the scale of the
daughter quark mass $m_s$ in the Cabibbo favored semileptonic
transition. The $D^+$ and $D^0$ semileptonic widths are equal
within the 3\% accuracy of the measurements, and the compositions
of their inclusive spectra are dominated by the lowest lying
resonances \cite{Adam:2006nu}. This result is explained by the
observation that the $s$ quark in the final state is usually
produced with a small enough momentum to be bound to the spectator
anti-quark in an $l=0$ $s\bar{q}$ meson \cite{Witherell:1993nt}.
$\Ds$ semileptonic decays share these kinematic features, and thus
an absolute measurement of $\Ds$ semileptonic decays sheds some
light also on inclusive processes. Specific decays contribute
valuable information on light meson properties. For example, the
fraction of semileptonic decays going into $\eta$ and $\etap$ is
sensitive to the pseudoscalar mixing angle, and may indeed shed
some light on $\eta-\etap$-glueball mixing
\cite{Anisovich:1997dz}. In addition, decays including $\pi\pi$
and $KK$ in the final state can elucidate the nature of exotic
light scalar mesons \cite{Dosch:2002hc,Fariborz:2008qi}.

$\Ds$ exclusive semileptonic decays have been studied by ARGUS,
CLEO, BaBar, and fixed target experiments. No absolute
measurements of branching fractions exist. The branching fraction
$\Ds^+\to \phi\ell^+\nu _\ell$, which is the most widely studied
\cite{link02j, butler94, frabetti93g, albrecht91,alexander90b}, is
generally normalized with respect to the decay $\Ds\to\phi\pi$.
However, the Dalitz plot for this mode shows the presence of a
significant broad scalar resonance whose contribution to the
observed yields changes depending upon the selection criteria
\cite{ds-had}. For this reason, this mode is not suitable for
normalization. Recently, the BaBar collaboration
\cite{babar:preprint} used the normalization mode $\Ds\to KK\pi$
with a mass cut of $\pm$ 10 MeV around the nominal $\phi$ as
suggested in Ref.~\cite{ds-had}, to obtain ${\cal B}(\Dsp\to \phi
e^+\nu _e)=(2.61\pm 0.03\pm 0.08\pm 0.15)$\%. CLEO measured
\cite{brandenburg} the ratio $[\Gamma(\eta\ell^+\nu
_{\ell})+\Gamma(\etap\ell^+\nu_\ell)]/[\Gamma(\phi \ell^+\nu
_\ell)]=1.67\pm 0.17\pm 0.17$. Finally, BES~\cite{bes:prd}
reported the inclusive branching fraction ${\cal B}(\Dsp\to e^+
\rm anything)=(7.7 ^{+5.7+2.4}_{-4.3-2.1})\%$. The uncertainties
are too large to allow a meaningful comparison between inclusive
and exclusive channels.

We use a data sample of $310$ pb$^{-1}$, collected at a
center-of-mass (CM) energy close to 4170 MeV, with the CLEO-c
detector \cite{cleoiii,cleoiiidr}. The momenta and directions of
charged particles are reconstructed in the tracking system, which
also provides charged particle identification information  based
on specific ionization ($dE/dx$). A Ring Imaging Cherenkov
Detector (RICH) completes the charged particle identification
system \cite{rich}, and is critical near 1 GeV, where the specific
ionization bands of the $K$ and $\pi$ overlap. The photon energy
and direction are measured in the CsI electromagnetic calorimeter,
whose energy measurement $E$, combined with the momentum $p$
information from the tracking system, provides the key electron
identification variable $E/p$. The CsI calorimeter measures the
electron and photon energies with an r.m.s. resolution of 2.2\% at
$E=1$~GeV and 5\% at $E=100$~MeV.

At $E_{\text{CM}}=4170$ MeV, the cross section for $\epm$
annihilation into $\Dsstarp\Dsm + \Dsp\Dsstarm$ is approximately
$0.9$ nb, while other charm production totals $\sim 7$ pb, and the
light quark continuum cross section is $\sim 12$ nb \cite{scan}.
We look for semileptonic decays of the $\Dsp$ in events in which
the $\Dsm$ is fully reconstructed in a hadronic mode (tagged
events). Each event also must include at least one isolated
photon, as either the $\Dsp$ or $\Dsm$ originates from a
$\Dsstar$. Here and throughout the paper, charge conjugate decays
are implied. Charged tracks are used to form the $\Dsm$ if their
fitted helical trajectory approaches the event origin within a
distance of $5$ mm in the azimuthal projection and $5$ cm in the
polar projection ($\theta$), where the azimuthal projection is in
the bend view of the solenoidal magnet. In addition, each track
must possess at least 50\% of the hits expected, must be within
the fiducial volume of the drift chamber, and must have a momentum
of at least 40 MeV. Pions and kaons are identified using $dE/dx$
and RICH if their momenta are above 700 MeV, otherwise only
$dE/dx$ identification is used. We form $\eta$ and $\pi ^0$
candidates from pairs of photons that deposit energy in the
calorimeter in a manner consistent with an electromagnetic shower
and are not matched to tracks. We require that the two photons
to have less than $3\sigma$ pull mass which is defined as the standard
deviation from the expected $\piz$ or $\eta$ mass. In the best
calorimeter region ($|\cos{\theta}|<$ 0.71) we use photons with
energies greater than 30 MeV, while in the endcap region
($0.93>|\cos{\theta}|>0.85$) we require an energy greater than 50
MeV.

The tag modes used are listed in Table I. Some tag specific
selection criteria are applied. For the modes  $\Dsm\to
\Kp\Km\pim$ and $\Dsm\to \Kp\Km\pim\piz$, the $\pi$ is required to
have a momentum greater than 100 MeV to suppress the background
from $\Dstar$ decays. Similarly, for the mode $\Dsm\to
\pip\pim\pim$, two pions of opposite charge must have momentum
greater than 100 MeV. The Charged track pairs used to reconstruct
$\Ks$ (via $\Ks\to\pip\pim$) are required to have an invariant
mass within 3$\sigma$ of the $\Ks$ mass. In addition, for  the
$\Dsm\to \Ks\Km$ and $\Dsm\to$\ksks\ tags, we require candidate
$\Ks$ to originate from a vertex displaced from the interaction
point, and the $\Ks$ momentum vector, obtained from a kinematic
fit of the charged $\pi$ momenta, must point back to the beam
spot. For resonance decays we select intervals in invariant mass
centered on the resonance masses \cite{pdg} and within $\pm$ 150
MeV for $\rhom\to \pim\piz$, $\pm$ 100 MeV for $K^\ast\to K\pi$,
$\pm$ 10 MeV for $\etap\to \eta \pip\pim$ or $\etap\to\rho\gamma$.
In addition, for the latter $\etap$ decay mode, we apply a
helicity angle cut which is the angle measured in the rest frame
of the decaying parent particle between the direction of the decay
daughter and the direction of the grandparent particle as
$|\cos{\theta_\pi}| <$ 0.8. Tags are required to have momentum
consistent with coming from $\Ds\Dsstar$ decay.
\begin{table}[htpb]
\begin{center}
\caption{Tagging modes and number of signal and background (Bkg)
events, determined from two-Gaussian fits to the invariant mass
distributions. The signal window is $\pm 2.5\sigma$ of the $\Dsm$
mass for all modes, except $\eta \rho^-$ ($\pm 2\sigma$). The last
two columns show the corresponding estimates of the number signal
and background tags accompanied by a reconstructed $\gamma$ from
$\Dsstar\to\gamma\Ds$ transition, which are determined from the
$\MMtwos$ spectrum (see text). }\label{tab:ntag}
\begin{tabular}{lcccc}
\hline \hline
  Mode & \multicolumn{2}{c}{Invariant Mass} & \multicolumn{2}{c}{$\rm MM^{\star 2}$}  \\
  ~~ & Signal & Bkg & Signal & Bkg \\
  \hline
  $K^{+}K^{-}\pi^{-}$   &13952 $\pm$ 232     & 11280 & 8245 $\pm$ 245 & 13970\\
  $K^{0}_{S}K^{-}$      & 2943 $\pm$ 128     &   561 & 1749 $\pm$ 146 &  1555\\
  $\eta \pi^{-}$        & 1806 $\pm$ 120     &  4747 & 1241 $\pm$ 123 &  3936\\
  \etappi               & 1231 $\pm$  55     &   415 &  907 $\pm$ 109 &  1036\\
  \kkpipiz              & 5300 $\pm$ 401     & 34419 & 2913 $\pm$ 289 & 24985\\
  $\pi^{+} \pi^{-} \pi^{-}$ & 4331 $\pm$ 716 & 25824 & 2439 $\pm$ 558 & 16619\\
  $K^{*-}{K}^{*0}$ & 1565 $\pm$ 114 &  1442 &  841 $\pm$  87 &  2440\\
  $\eta\rho^{-}$            & 4002 $\pm$ 254 & 22044 & 2168 $\pm$ 268 & 18450\\
  \epirho               &   2515 $\pm$  342  & 18593 & 1817 $\pm$ 212 & 12061\\
  \hline
  Sum                      & 37645 $\pm$ 978 &119325 &22320 $\pm$ 792 & 95052\\
  \hline \hline
\end{tabular}
\end{center}
\end{table}

We further select tags using the recoiling mass $M_{\text{rec}}$,
\begin{eqnarray}
M_{\text{rec}} = \sqrt{(E_{\text{CM}} - E_{\Dsm})^2 -
(\mathbf{p}_{\text{CM}}-\mathbf{p}_{\Dsm})^2 },
\end{eqnarray}
where $E_{\text{CM}}$ ($\rm \mathbf{p}_{CM})$ is the CM energy
(momentum), $E_{\Ds}$ ($\mathbf{p}_{\Ds}$) is the tag energy
(momentum). $D_s^\star$ daughter tags peak broadly in
$M_{\text{rec}}$ due to the presence of the photon in the tag side,
while directly produced $\Dsm$ tags have a narrow peak. We accept
events for which $M_{\text{rec}}$ is within $\text{-55\ MeV} \leq
M_{\text{rec}} - M_{D_s^\star} \rm < 55\ MeV$. This cut is broad
enough to encompass both the narrow peak associated with $\Ds$ tags
and almost all the broad component associated with tags that originate from
$\Dsstar$. Then we fit the invariant mass distribution $\MDs$ of
these events, using a two-Gaussian shape for the signal plus a
polynomial background shape. The signal component allows us to
define an effective $\sigma =f_1\sigma _1 +(1-f_1)\sigma _2$ where
$\sigma _1$ and $\sigma _2$ are the standard deviations of the two
Gaussian components and
 $f_1$ is the fractional area of the first Gaussian. We require that the candidate invariant
mass to be within $2.5\sigma$ ($2\sigma$ for the $\eta\rho$ mode)
of the nominal $\Ds$ mass \cite{pdg}. Random $\Ds$ backgrounds are
estimated through sideband samples. We then combine the tag with a
well reconstructed $\gamma$ and calculate the missing mass squared
$\MMtwos$,  the square of the invariant mass of the system recoiling against the
$\gamma$-tag pair.
\begin{equation}
\MMtwos = (E_{\text{CM}}-E_{\Dsm}-E_{\gamma})^2-
(\mathbf{p}_{\text{CM}}-\mathbf{p}_{\Dsm}-\mathbf{p}_{\gamma})^2,
\end{equation}
where $E_{\gamma}$ ($\mathbf{p}_{\gamma}$) is the energy
(momentum) of the additional $\gamma$. In order to improve the
$\MMtwos$ resolution, we use a kinematic fit that constrains the
$\Ds$ decay products to $\MDs$ and conserves overall momentum and
energy.

\begin{figure}[htbp]
\epsfig{figure=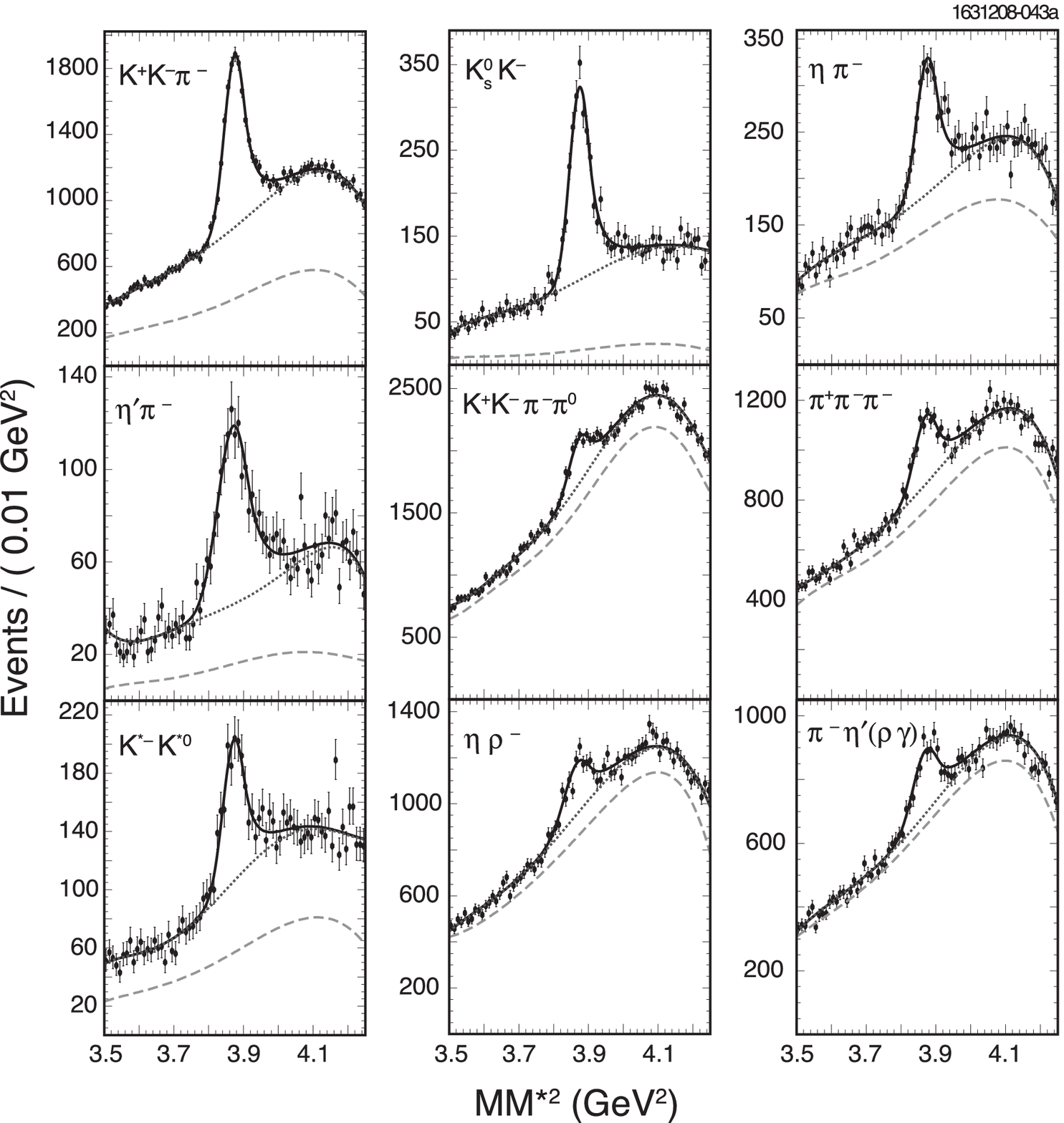,width=3.5 in}
\caption{The $\MMtwos$ distribution from events with a photon in
addition to the $\Dsm$ tag.  In each plot, the $\Dsm$ mode studied is indicated, and the solid curve represents a fit to a Crystal Ball function (signal), the dotted curve represents the total background, and the dashed curve represents the background composed of random $\Ds$ tags, constrained by the sideband sample. Both terms are well
described by 5$^{\rm th}$ order Chebychev polynomial background functions.}\label{m2s:fig}
\end{figure}

Fig.~\ref{m2s:fig} shows the $\MMtwos$ distribution for the nine
tags considered.  In order to estimate the number of tags used for
further analysis, we use  a two-dimensional binned maximum
likelihood fit of the measured $\MMtwos$ and $\MDs$ distributions.
We consider three components in the fit: a signal, comprising true
tags accompanied by the photon from the $\Dsstar$, a background
composed by true tags combined with a random photon, and a second
background comprising false tags. We infer the $\Ds$ combinatoric
background from two 5$\sigma$ (4$\sigma$ for the $\eta\rho$ mode)
wide intervals on both sides of the $\MDs$ signal peak. The
$\MMtwos$ signal fit is improved by using a probability
distribution function (PDF) derived from fully reconstructed
$\Ds\Dsstar$ events. We fit signal and sideband intervals in
$\MDs$ simultaneously. The sideband intervals constrain the shape
of the random $\Ds$ background. We then extract the tag yield from
the fitted signal function integrated within $\pm$ 2.5$\sigma$
around the $\MMtwos$ most probable value. For the nine modes
considered, Table~\ref{tab:ntag} shows the number of signal and
background tags, as well as the signal and background tags
reconstructed in conjunction with an isolated photon .

We next describe reconstruction of the semileptonic decays. For
any given tag-photon combination, we seek a candidate $e^+$ and a
set of hadrons. Positrons are identified on the basis of a
likelihood ratio constructed from three inputs: the ratio between
the energy deposited in the calorimeter and the momentum measured
in the tracking system, the specific ionization $dE/dx$ measured
in the drift chamber, and RICH information \cite{d0excl}. Our
selection efficiency averages 0.95 in the momentum region 0.3-1.0
GeV, and 0.71 in the region 0.2-0.3 GeV. The average fractions of
charged $\pi$ and $K$ incorrectly identified as positrons averaged
over the relevant momentum range are approximately 0.1\%. We study
events containing $\phi$, $\eta$, $\etap$, $\Ks$, $\Kstar$, and
$f_0$ in the final state. Track and $\gamma$ selection criteria,
as well as resonance cuts are the same as used in the tag
reconstruction except that we select candidates with invariant masses
 within $\pm$ 10
MeV  of the known $\phi$ mass \cite{pdg} for $\phi\to \Kp\Km$, and within $\pm$ 75 MeV of the known $\Kstar$ mass for $\Kstar\to \Kp\pim$.
Among the $\eta$ candidates not used in forming a tag, we choose
the one with the smallest pull mass to form $\eta e^+ \nu_e$ and
$\etap e^+ \nu_e$ candidates. We use $\etap\to \eta\pip\pim$ only.
For the channel $\Kstarz\to\Kp\pim$, the $K$ and $e$ must have the
same charge. Finally, we form $f_0$ candidates by combining two
pions of opposite charge and require that their invariant mass be
within 100 MeV of the known $f_0$ mass. In each case we require
that the event have no unused tracks, and that the tag and
semileptonic candidate have opposite charge.

\begin{figure}[hbt]
\epsfig{figure=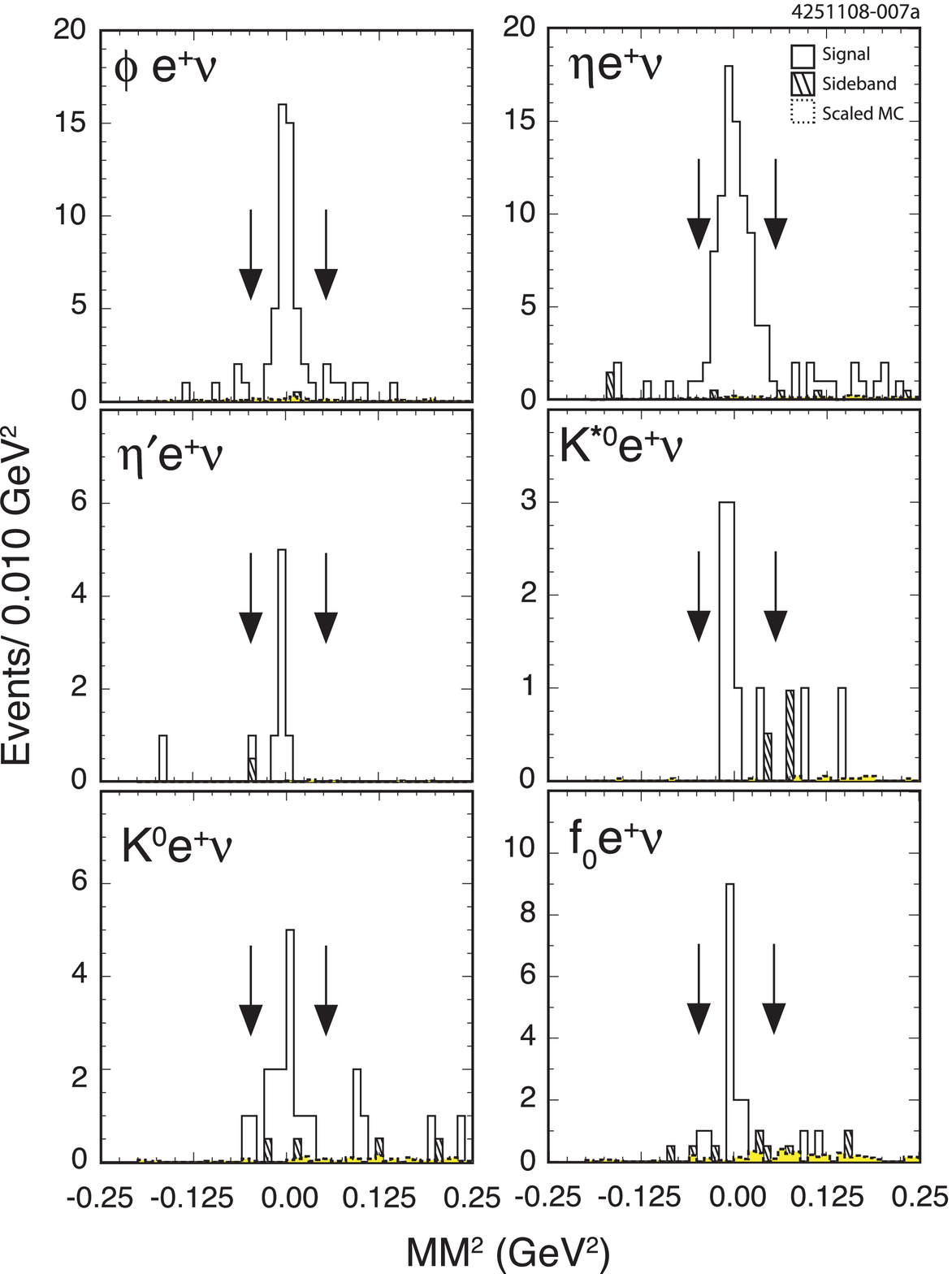,height=4.55in}
\caption{The $\rm MM^2$ distribution for tagged semileptonic
events in the exclusive modes: \phienup, \etaenup, \etapenup,
\kstenup, \kzeenup, and \fzeenup. The solid entries correspond to
the events from signal regions, the hatched entries are the events
from sideband and the dashed entries represent the scaled
background events from generic MC.} \label{fig:mm2}
\end{figure}

For each $\gamma$ candidate, we perform two kinematic fits, one
assuming that the $\gamma$ combines with the tag to form a
$\Dsstarm$, the other assuming that the semileptonic decay comes
from a $\Dsstarp$ parent. We require the $\Ds\Dsstar$ pair to
conserve energy and momentum in the CM frame, and the mass of the
candidate $\Ds$ to be consistent with the known mass. When we
assume the tag to be the daughter of a $\Dsstarm$, we constrain
the energy of the photon plus tag candidate to be consistent with
the expected $\Dsstarm$ energy, otherwise we constrain the energy
of the tag candidate to be consistent with the $\Dsm$ energy in
the CM system. Finally we choose the photon and hypothesis with
the smallest $\chisq$ and calculate the missing mass squared $\rm
MM^2$ defined as
\begin{eqnarray}
\text{MM}^2=
(E^{\ast}_{\text{CM}}-E^{\ast}_{D_S^-}-E^{\ast}_{\gamma}-E^{\ast}_e-E^{\ast}_{\text{had}})^2
\\ \nonumber
-(-\mathbf{p}^{\ast}_{\Dsm}-\mathbf{p}^{\ast}_\gamma-\mathbf{p}^{\ast}_e-\mathbf{p}^{\ast}_{\text{had}})^2,
\end{eqnarray}
where $E^{\ast}_e$ ($\mathbf{p}^{\ast}_e$) is the energy (momentum)
of the positron candidate and $E^{\ast}_{\text{had}}$ ($\rm
\mathbf{p}^{\ast}_{\text{had}}$) is the energy (momentum) of the
hadron candidate in the CM system. For signal events, $\rm MM^2$ is
the $\nu _e$ invariant mass squared and thus it peaks at zero.
Fig.~\ref{fig:mm2} shows the measured $\rm MM^2$ for each final
state summed over all tag modes used. We require signal events to
have a $\rm |MM^2| < 0.05$ GeV$^2$. We estimate the background
coming from random $\Dsm$ tags by studying the sideband samples, and
the remaining background by studying a sample of simulated
$D\bar{D}$ events that is 20 times bigger than our data set.
Fig.~\ref{fig:mm2} shows the signal and background distributions for
the six modes studied. The $MM^2$ shapes are well modeled by the
signal Monte Carlo simulation. We show also the background estimates
from the sideband data sample, as well as the background predictions from a 
Monte Carlo simulation of charm meson decays at this center-of-mass energy. Note that the background is small in
all the modes considered. We have also investigated backgrounds
produced by random photons associated with a true semileptonic
event, and we found these to be even smaller, thus we do not
subtract them.

We evaluate exclusive branching fractions for semileptonic decays including the hadron $i$ through the relationship
\begin{equation}
{\cal B}_i \equiv \frac{\Sigma_{\alpha} (n^i_{\alpha}-n^i_{\alpha, \text{bkg}})}{\epsilon^i _{\text{SL}}(\Sigma _{\alpha} n_{\alpha} ){\cal B}_i^{\text{had}}}\label{eq:bfdz}
\end{equation}
where the index $\alpha$ runs over the tag modes, the index $i$
represents a specific hadronic final state, ${\cal
B}_i^{\text{had}}$ identifies the branching fraction for that
state, and $\epsilon^i _{\text{SL}}$ represents the average
efficiency for finding the exclusive semileptonic decay in the tag
sample used. We evaluate the semileptonic efficiency by
considering two Monte Carlo (MC) samples. The first contains a tag
event accompanied by a semileptonic decay (double-tag sample),
while the latter contains a tag accompanied by a generic $\Ds$
decay (single-tag sample). The ratio of the single and double tag
efficiencies is the desired $\rm \epsilon _{SL}$ for each tag.
Table~\ref{tab:sl} shows the signal, and background yields, $\rm
\epsilon _{SL}$ and the branching fractions determined for the six
semileptonic channels. We derive $\rm \epsilon _{SL}$ from each
tag mode independently, and then we compute their average weighted
by tag abundance. The efficiency obtained with this method
accounts for the different tag efficiency in semileptonic and
generic $\Ds$ decays. We treat the exclusive channel $\phi
e^+\nu_e$ slightly differently: as the $\phi$ reconstruction
efficiency is strongly momentum dependent, we perform the analysis
in five 200 MeV wide momentum bins. We attribute all the signal
events in the $\Kp\Km e^+\nu _e$ in the final state to form the
$\phi e \nu_e$ channel. Recent BaBar studies \cite{babar:preprint}
have estimated the $S$-wave fraction to be
$(0.22^{+0.12}_{-0.08})$ \% of the decay rate, smaller than our
statistical uncertainty in our $\phi e \nu_e$ branching fraction.
We also look for $\pi\pi$, $KK$, and $K\pi$ outside the mass
windows corresponding to the resonant states studied and we found
no evidence for additional channels.

\begin{table}[hbt]
\begin{center}
\caption{The signal and background yields, the semileptonic
efficiency $\epsilon^{i}_{\text{SL}}$ and the derived branching
fractions for the six semileptonic channels studied. The \fzeenup\
branching fraction quoted represents the product branching
fraction ${ \cal B}(\Dsp\to f_0 e^+\nu_e) \times {\cal B}(f_0 \to
\pip\pim)$, which is the dominant decay mode in
Ref.~\cite{pdg}.}\label{tab:sl}
\begin{tabular}{lcccc}
  \hline \hline
Signal Mode & $n^{i}$ & $n^{i}_{\text{bkg}}$ & $\epsilon^{i}_{\text{SL}}(\%)$ & ${\cal B}(\%)$\\
  \hline
\phienup & 45.50 & 0.06 & 17.79 $\pm$ 0.33 & 2.29 $\pm$ 0.37\\
\etaenup & 82.49 & 0.32 & 37.65 $\pm$ 0.27 & 2.48 $\pm$ 0.29\\
\etapenup&  7.50 & 0.06 & 21.04 $\pm$ 0.22 & 0.91 $\pm$ 0.33\\
\kzeenup & 13.99 & 0.29 & 33.14 $\pm$ 0.26 & 0.37 $\pm$ 0.10\\
\kstenup &  7.50 & 0.18 & 27.52 $\pm$ 0.23 & 0.18 $\pm$ 0.07\\
\fzeenup & 13.99 & 0.88 & 46.79 $\pm$ 0.31 & 0.13 $\pm$ 0.04\\

\hline \hline
\end{tabular}
\end{center}
\end{table}

We consider several sources of systematic uncertainty. The
dominant component is associated with the number of tags, which is
affected by the lack of our knowledge on the random $\gamma$ in
the background PDFs. We estimate it by repeating the fit with a
variety of shapes, namely polynomials of different order, or
special shapes derived from MC simulation, and obtain an
uncertainty of 3.6\%. Systematic uncertainties associated with
hadron selection such as tracking (0.3\% per charged particle),
$K$ and $\pi$ identification (0.6\% and 0.3\% respectively), and
$\eta$ selection criteria (2\%) have been studied extensively
\cite{dhad}. Similarly, the $\Ks$ selection criteria are derived
from Ref.~\cite{kseff}, and have (0.8\%) uncertainty. The
systematic uncertainty on the electron identification efficiency
(1\%) is assessed by comparing radiative Bhabha samples, Bhabha
events embedded in hadronic events, and MC samples. The
requirements that there are no extra tracks in the event and that
the net charge is zero have been evaluated with a data sample
comprised of two hadronic tags. The comparison between results
obtained with this sample and corresponding MC samples give an
overall systematic uncertainty of 0.6\% from these two
requirements. Finally, we consider the dependence of the
efficiency for semileptonic decays on the form factors. The CLEO
MC uses the form factors predicted by the ISGW2 model
\cite{ISGW2}. We have generated also samples based on simple pole
form factors and compared the efficiencies derived with the two
methods to estimate this effect. The related systematic
uncertainty ranges from 0.1\% to 2.4\%.

We check the normalization of our branching fractions by measuring
the well known branching fraction ${\cal B}(D^0\to K^{-} e^+\nu
_e)$ using a $\Dm\Dstarp$ sample from the same data set. We
reconstruct $DD^\star\to D^-\Dstarp\to D^-\pip \Dz$ decays, where
the $D^0$ decays into $K^- e^+\nu _e$, and the $\Dm$ decays into
these six hadronic exclusive final states: $D^{-} \to$ \kpipim,
$D^{-} \to$ \kpipipizm, $D^{-} \to$ \kzpim, $D^{-} \to$ \kzpipizm,
$D^{-} \to$ \kzpipipim, $D^{-} \to$ \kkpipm. The selection
criteria and analysis procedure are the same as used in
reconstructing the $\Ds$ semileptonic decays. We get in total
14759 $\pm$ 203 of tagged events and 350 $\pm$ 18 signal events,
and using Eq.~(\ref{eq:bfdz}), we derive a branching fraction
${\cal B}(D^0\to K^- e^+ \nu _e)=(3.45 \pm
 0.21)\%$.

This result is in agreement with the two most recent absolute
measurements: ${\cal B}(D^0\to K^- e^+ \nu _e) = (3.61 \pm 0.05
\pm 0.05)\%$ from CLEO-c~\cite{dz:cleo}, based on 281 pb$^{-1}$
data at the $\psi(3770)$, and ${\cal B}(D^0\to K^- e^+ \nu _e) =
(3.45 \pm 0.07 \pm 0.20)\%$ from Belle~\cite{dz:bell}.

All the measurements reported here are first absolute measurements
of exclusive semileptonic $\Ds$ decays, moreover this is the first
report of  Cabibbo suppressed final states and scalar meson above the
threshold to decay in the observed final state. For the six $\Ds$ semileptonic decays considered,
Table~\ref{tab:br} shows the derived branching fractions including
the systematic errors.

\begin{table}[hbt]
\begin{center}
\caption{The derived branching fractions including the systematic
errors for the six semileptonic channels studied. The \fzeenup\
branching fraction quoted represents the product branching
fraction ${ \cal B}(\Dsp\to f_0 e^+\nu_e) \times {\cal B}(f_0 \to
\pip\pim)$, which is the dominant decay mode in
Ref.~\cite{pdg}.}\label{tab:br}
\begin{tabular}{lc}
  \hline \hline
  Signal Mode &  ${\cal B}(\%)$\\
  \hline
$\Dsp\to\phi e^+\nu_e$ & $ 2.29 \pm 0.37 \pm 0.11 $ \\
$\Dsp\to\eta e^+\nu_e$ & $ 2.48 \pm 0.29 \pm 0.13 $ \\
$\Dsp\to\etap e^+\nu_e$& $ 0.91 \pm 0.33 \pm 0.05 $ \\
$\Dsp\to K^0 e^+\nu_e$ & $ 0.37 \pm 0.10 \pm 0.02 $ \\
$\Dsp\to K^{\star 0} e^+\nu_e$ &  $ 0.18 \pm 0.07 \pm
0.01 $ \\
$\Dsp\to f_0 e^+\nu_e$ & $ 0.13 \pm 0.04 \pm 0.01 $ \\
 \hline \hline
\end{tabular}
\end{center}
\end{table}

These results allow us to draw several interesting conclusions.
 The sum of the branching fractions measured
imply ${\cal B}(\Dsp\to X e^+\nu _e)=(6.47\pm 0.60)\%$, about 16
\% below the value ($7.82 \pm 0.13$)\%, inferred from 
measured $\Dp$ and $Dz$ inclusive semileptonic branching fraction \cite{dincl} and the charmed
meson lifetimes \cite{pdg}. We
have searched for additional hadronic final states formed
with two charged tracks, as well as from two charged tracks and a
$\pi^0$, and found no evidence for semileptonic decays including
other hadronic final states. No other significant branching
fraction is expected. This result is
consistent with the predictions of the ISGW2 model \cite{ISGW2},
supporting the conjecture that SU(3) is broken in charm
semileptonic decays. On the other hand, the difference in widths
may arise from non factorizable contributions at the level of
$\sim$ 10\% \cite{voloshin}. The ratio ${\cal B}(\Dsp\to \etap
e^+\nu _e)/{\cal B}(\Dsp\to \eta e^+ \nu _e)=0.36\pm 0.14$, is in
agreement with the previous CLEO result \cite{brandenburg}. The
ISGW2 model involves a $\eta/\etap$ mixing angle close to
$-10^\circ$, which is the minimum value obtained from mass
formulae \cite{pdg} if a quadratic approximation is used.
According to Ref.~\cite{Anisovich:1997dz}, the measured ratio is
consistent with a pseudoscalar mixing angle of about $-17^\circ$,
provided that a glueball component probability of the order of
10\% is present in the $\etap$. Finally, we have the first
measurement of a $\Ds$ semileptonic decay including a scalar meson
above the threshold for decay to the observed final state, which opens up the exciting possibility of
elucidating the nature of exotic light mesons \cite{Dosch:2002hc}.

\section{Acknowledgements}
We gratefully acknowledge the effort of the CESR staff in providing us with excellent luminosity and running conditions. This work was supported by the National Science Foundation and the U.S.
Department of Energy.


\begin{thebibliography}{00}
\bibitem{voloshin}
 M.~B.~Voloshin,
  Phys.\ Lett.\  B {\bf 515}, 74 (2001).



\bibitem{ISGW2}
  D.~Scora and N.~Isgur,
  Phys.\ Rev.\  D {\bf 52}, 2783 (1995).

\bibitem{Adam:2006nu}
  N.~E.~Adam {\it et al.}  (CLEO Collaboration),
  Phys.\ Rev.\ Lett.\  {\bf 97}, 251801 (2006).

\bibitem{Witherell:1993nt}
  M.~S.~Witherell,
  AIP Conf.\ Proc.\  {\bf 302}, 198 (1994).

\bibitem{Anisovich:1997dz}
  V.~V.~Anisovich, D.~V.~Bugg, D.~I.~Melikhov and V.~A.~Nikonov,
  Phys.\ Lett.\  B {\bf 404}, 166 (1997).

\bibitem{Dosch:2002hc}
  H.~G.~Dosch and S.~Narison,
  Nucl.\ Phys.\ Proc.\ Suppl.\  {\bf 121}, 114 (2003).


\bibitem{Fariborz:2008qi}
  A.~H.~Fariborz, R.~Jora and J.~Schechter,
  arXiv:0810.4640 [hep-ph].


\bibitem{link02j} J.M. Link {\it et al.}, (FOCUS Collaboration), Phys. Lett. B. {\bf 541}, 243 (2002).

\bibitem{butler94} F. Butler {\it et al.}  (CLEO Collaboration), Phys. Lett. B. {\bf 325}, 255 (1994).

\bibitem{frabetti93g}
  P.~L.~Frabetti, {\it et al.}, Phys. Lett. B. {\bf 313}, 253 (1993).

\bibitem{albrecht91} H. Albrecht {\it et al.}  (ARGUS Collaboration), Phys. Lett. B. {\bf 245}, 315 (1990).


\bibitem{alexander90b}
  J.~P.~Alexander {\it et al.}  (CLEO Collaboration),
  Phys.\ Rev.\ Lett.\  {\bf 65}, 1531 (1990).

\bibitem{ds-had}
  J.~P.~Alexander {\it et al.}  (CLEO Collaboration),
  Phys.\ Rev.\ Lett.\  {\bf 100}, 161804 (2008).

\bibitem{babar:preprint} B. Aubert {\it et al.} (BaBaR Collaboration),
  Phys.\ Rev.\  D {\bf 78}, 051101 (2008).

\bibitem{brandenburg}
  G.~Brandenburg {\it et al.}  (CLEO Collaboration),
  Phys.\ Rev.\ Lett.\  {\bf 75}, 3804 (1995).

\bibitem{bes:prd}
  J.~Z.~Bai {\it et al.}  (BES Collaboration),
  Phys.\ Rev.\  D {\bf 56}, 3779 (1997).

\bibitem{cleoiii} Y. Kubota {\it et al.}, Nucl.\ Instrum.\ Meth.\ A {\bf 320}, 66 (1992).

\bibitem{cleoiiidr}
      D.~Peterson {\it et al.}, Nucl.~Instrum.~Methods Phys.~Res., Sec.~A
      {\bf 478}, 142 (2002).

\bibitem{rich} M.~Artuso {\it et al.}, 
    Nucl.\ Instrum.\ Meth.\ A {\bf 502}, 91 (2003). 

\bibitem{scan} D. Cronin-Hennessy {\it et al.} (CLEO Collaboration),
arXiv:0801.3418.

\bibitem{pdg} W. M. Yao {\it et al.} (Particle Data Group), Journal of Physics, {\bf G 33}, 1(2006).


\bibitem{d0excl}
 T.~E.~Coan {\it et al.}  (CLEO Collaboration),
  Phys.\ Rev.\ Lett.\  {\bf 95}, 181802 (2005).






\bibitem{dhad}S. Dobbs {\em et al.} (CLEO Collaboration), Phys. Rev D {\bf 76}, 112001 (2007).


\bibitem{kseff} J.~L.~Rosner {\it et al.} (CLEO Collaboration), Phys.\ Rev.\ Lett.\  {\bf 100}, 221801 (2008).

\bibitem{dz:cleo} J. Y. Ge {\em et al.} (CLEO Collaboration), arXiv:0810.3878 [hep-ex] (2008).

\bibitem{dz:bell} L. Widhalm, Phys. Rev. Lett. {\bf 97}, 061804 (2006).

\bibitem{dincl}
 N.~E.~Adam {\it et al.}  [CLEO Collaboration],
  Phys.\ Rev.\ Lett.\  {\bf 97}, 251801 (2006)
  [arXiv:hep-ex/0604044].







\end{thebibliography}
\end{document}